\documentclass[twocolumn,aps,prl,superscriptaddress,showpacs]{revtex4-1}
\usepackage[dvips]{graphicx,psfrag}
\usepackage{amssymb,amsfonts,amsmath}
\usepackage{epsfig}
\scrollmode

\begin{document}

\title{Optimal fits of diffusion constants from single time data points of Brownian trajectories}

\author{Denis Boyer}
\email{boyer@fisica.unam.mx}
\affiliation{Instituto de F\'{\i}sica, Universidad Nacional Aut\'onoma de M\'exico,
D.F. 04510, M\'exico}
\author{David S. Dean}
\email{david.dean@u-bordeaux1.fr}
\affiliation{Universit\'e de  Bordeaux and CNRS, Laboratoire Ondes et
Mati\`ere d'Aquitaine (LOMA), UMR 5798, F-33400 Talence, France}
\author{Carlos Mej\'{\i}a-Monasterio}
\email{carlos.mejia@upm.es}
\affiliation{Laboratory of Physical Properties, Technical University
of Madrid, Av. Complutense s/n 28040, Madrid, Spain}
\affiliation{Department of Mathematics and Statistics,
 P.O.  Box 68 FIN-00014, Helsinki, Finland}
\author{Gleb Oshanin}
\email{oshanin@lptmc.jussieu.fr}
\affiliation{Laboratoire de Physique Th\'eorique de la Mati\`ere
Condens\'ee (UMR CNRS 7600), Universit\'e Pierre et Marie Curie, 4
place Jussieu, 75252 Paris Cedex 5 France}

\date{\today}

\begin{abstract}
  Experimental  methods based  on single  particle tracking  (SPT) are
  being increasingly employed in the physical and biological sciences,
  where  nanoscale  objects  are  visualized with  high  temporal  and
  spatial resolution.   SPT can probe interactions  between a particle
  and  its environment  but the  price to  be paid  is the  absence of
  ensemble  averaging and a  consequent lack  of statistics.   Here we
  address the  {\em benchmark} question  of how to  accurately extract
  the diffusion  constant of {\em one} single  Brownian trajectory. We
  analyze a class  of estimators based on weighted  functionals of the
  square displacement.   For a certain  choice of the  weight function
  these  functionals  provide  the  true ensemble  averaged  diffusion
  coefficient,  with a  precision that  increases with  the trajectory
  resolution.
\end{abstract}

\pacs{05.40.Jc, 31.15.xk, 87.16.dp, 61.43.Er} \maketitle


Single particle tracking (SPT) generates the time series of the
position of an individual particle trajectory ${\bf B}_t$ in a medium
(see, e.g., \cite{bra,saxton}).  Properly interpreted, the information
so obtained provides an insight into the mechanisms and forces that
drive or constrain the motion of the particle \cite{moerner}.
Nowadays single particle tracking is extensively used to characterize
the microscopic rheological properties of complex media \cite{mason}
and to probe the active motion of biomolecular motors
\cite{greenleaf}. In biological cells and complex fluids, SPT methods
have become instrumental in demonstrating deviations from standard
Brownian motion (BM)
\cite{golding,weber,bronstein,seisenberger,weigel}.

The reliability of the information drawn from SPT analysis is not
always clear: data is obtained at high temporal and spatial resolution
but at the expense of statistical sample size.  Time averaged
quantities associated with a given trajectory are subject to large
fluctuations across trajectories.  For a wide class of anomalous
diffusion problems, for instance, time-averages of certain particle's
observables are, by their very nature, random variables distinct from
their ensemble averages \cite{rebenshtok,ralf,he,lubelski}.

Even though standard BM is much better understood than anomalous
diffusion processes, averaging problems persist and complicate the
analysis of single trajectories. Moreover, in bounded systems,
substantial manifestations of sample-to-sample fluctuations occur in
first passage time phenomena \cite{carlos}.  Standard fitting
procedures applied to a finite Brownian trajectory unavoidably lead to
fluctuating estimates $D_f$ of the diffusion coefficient, due to
different thermal histories, particle interactions with different
defects, or simply due to blur and localization errors, as discussed
in \cite{berglund,michalet,mb}.  In fact, $D_f$ might be very
different from the true ensemble average value $D$, as noticed in SPT
measurements of diffusion along DNA \cite{austin}, in the plasma
membrane \cite{saxton} or in the cytoplasm of mammalian cells
\cite{goulian}.

The broad dispersion of estimate values extracted from common SPT
analysis raises an important question: Does an optimal methodology
able to determine the diffusion coefficient from just one single
trajectory exist?  Clearly, it is highly desirable to have an
estimator of this kind even for hypothetical pure cases, such as the
unconstrained standard BM with perfectly known location at a given
time.  Such an estimator should possess an ergodic property, {\em
  i.e.}, its most probable value should converge to the ensemble
average and its variance should vanish as the observation time
increases.  Besides, the knowledge of the distribution of a family of
estimators could provide a way to disentangle the effects of the
medium complexity or localization errors from variations due to the
thermal noise driving microscopic diffusion.

In this 
paper we study the ergodic properties of
weighted one-time 
fits to a BM trajectory.  We focus on a family of
weighted least-squares estimators ($u_{\mu}$) of the diffusion
coefficient of standard $d$-dimensional BM, given by the following
quadratic functionals of a trajectory ${\bf B}_t$:
\begin{equation}
\label{u}
u_{\mu} = \frac{A_{\mu}}{T} \int^{T}_{0} \, dt \, \omega(t) \,
{\bf B}^2_t \,,
\end{equation}
where ${\bf B}_{t=0}={\bf 0}$,  $\omega(t)$ is some \lq \lq trial" 
weight function of the form:
\begin{equation}
\label{weight1}
\omega(t) = \frac{1}{(t_0 + t)^{\mu}} \,,
\end{equation}
$\mu$ being a tunable exponent, $T$ - the observation time, $t_0$ - a
certain lag time ($t_0 \ll T$) and $A_{\mu}$ - the normalization
constant.  
The term \lq \lq least-squares" and the choice of the
weight function will be made clear below.

Here we evaluate the distribution $P(u_{\mu})$ for arbitrary $\mu$ and
spatial dimension $d$.  To easily compare the accuracy of estimators
with different values of $\mu$, we chose $A_{\mu}$ such that
$\mathbb{E}\left\{u_{\mu}\right\} \equiv 1$, where the symbol
$\mathbb{E}\{\ldots\}$ denotes the ensemble average.  Hence,
$u_{\mu}=D_f/D$ with $D$ being the diffusion constant,
\begin{equation}
\label{msd}
D = \frac{\mathbb{E}\left\{{\bf B}^2_t\right\}}{2 d t} \,,
\end{equation}
and $D_f$ - its estimate from ${\bf B}_t$. The best choice of $\mu$
should produce $P(u_{\mu})$ whose maximum $u^*$ is the closest to the
ensemble averaged value $1$ and have the smallest variance ${\rm
  Var}(u_{\mu})$. Ultimately, we seek the choice at which $u_{\mu}$ is
ergodic, {\it i. e.} $D_f \to D$, independently of ${\bf B}_t$ as
$\varepsilon \equiv t_0/T \to 0$.

Before we proceed, two remarks are in order.  First, note that $\mu
= - 1$ corresponds to the standard least square estimate (LSE) of the
square displacement  \cite{saxton,berglund,goulian,saxton2}.  
The case $\mu = 1$ arises when the unconditional probability of observing 
the whole trajectory ${\bf B}_t$ is maximized (assuming that it is
Brownian). It is the so-called maximum likelihood estimate (MLE), known 
to be more accurate than the LSE
\cite{berglund,michalet,mb,boyer,boyer1}.

We next give a physical interpretation of the estimators in
Eq.~(\ref{u}). Consider a least squares functional:
\begin{equation}
\label{func}
F = \frac{1}{2} \int^T_0 \frac{\omega(t) \, dt}{t} \,
\left({\bf B}^2_t - 2 d D_f t\right)^2 \,,
\end{equation}
which we generalize by adding a time dependent weight function
$\omega(t)$ (the standard choice -LSE- is $\omega(t) \equiv t$).  The
value of $D_f$ that minimizes $F$ is:
\begin{equation}
\frac{D_f}{D} = \left(\frac{1}{T} \int^T_0 dt \, \omega(t) \, {\bf B}^2_t\right)/
\left(\frac{2 d D}{T} \int^T_0 dt \, t \, \omega(t) \right)
\end{equation}
One recovers Eq.~(\ref{u}) by choosing the weight function $\omega(t)
= (t_0 + t)^{-\mu}$ and identifying the denominator with
$A_{\mu}$. Hence $u_{\mu}$ minimizes a functional (\ref{func}) and can
be referred to as a weighted least-squares estimator.

The moment generating function $\Phi(\sigma)$ of the random variable
$u_{\mu}$, Eq.~(\ref{u}), is defined as
\begin{equation}
\Phi(\sigma) = \mathbb{E}\left\{e^{- \sigma u_{\mu}}\right\}.
\end{equation}
This function can be calculated using the Feynman-Kac formula (see
Refs.\cite{boyer,boyer1} for more details).  For $\mu \neq 2$, we find
that to leading order in $\varepsilon = t_0/T$
\begin{equation}
\label{alpha<2}
\Phi(\sigma) = \left[\Gamma\left(\nu\right)
\left(\frac{\sigma}{\chi_1}\right)^{\frac{\nu - 1}{2}} {\rm I}_{1 - \nu}
\left(2 \sqrt{\frac{\sigma}{\chi_1}}\right)\right]^{-d/2} \,,
\end{equation}
for $\mu < 2$, while for $\mu > 2$ it obeys
\begin{equation}
\label{alpha>2}
\Phi(\sigma) = \left[\Gamma\left(1-\nu\right)
\left(\frac{\sigma}{\chi_2}\right)^{\frac{\nu}{2}} {\rm I}_{- \nu}
\left(2 \sqrt{\frac{\sigma}{\chi_2}}\right)\right]^{-d/2} \,,
\end{equation}
where $\nu = 1/(2 - \mu)$, $\chi_1 = d (2 - \mu)/2$, $\chi_2 = d (\mu
- 2)/2 (\mu - 1)$ and $I_{\mu}(z)$ is the modified Bessel function
\cite{abramowitz}.

\begin{figure}[ht]
  \psfrag{VAR}[b][b][1]{$\mathrm{Var}(u_\mu)$}
  \psfrag{A}[b][b][1]{$\mu$}
  \centerline{\includegraphics*[width=0.45\textwidth]{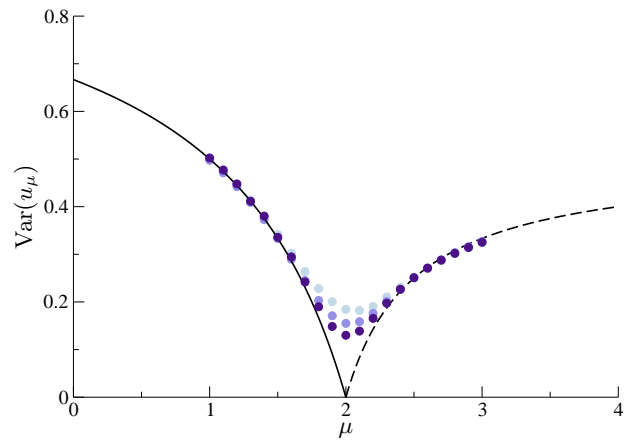}}
  \caption{(color online) The variance of the distribution
    $P(u_{\mu})$ for different values of $\mu$.  The curves correspond
    to Eq.~(\ref{varvar}).  The symbols correspond to the values
    obtained from direct numerical simulations of 3D random walks for
    (from light to dark) $\varepsilon = 5\times10^{-5}$,
    $5\times10^{-6}$ and $5\times10^{-7}$.}
\label{fig1}
\end{figure}

The variance ${\rm Var}(u_{\mu})$ of $P(u_{\mu})$ is obtained by
differentiating Eqs.~(\ref{alpha<2}) or (\ref{alpha>2}) twice with
respect to $\sigma$.  For arbitrary $\mu \neq 2$ it is then given
explicitly by
\begin{equation}
\label{varvar}
{\rm Var}(u_{\mu}) = \frac{2}{d}
\begin{cases} (2 - \mu)/(3 - \mu), & \text{$\mu < 2$,}
\\
(\mu - 2)/(2 \mu - 3), &\text{$\mu > 2$.}
\end{cases}
\end{equation}
The consequence of the latter equation is shown in Fig.~\ref{fig1}.
Unexpectedly, the variance can be made arbitrarily small at leading
order in $\varepsilon$ by taking $\mu$ gradually closer to $2$, either
from above or from below! The slopes at $\mu = 2^+$ and $\mu = 2^-$
appear to be the same, so that the accuracy of the estimator will be
the same for approaching $\mu = 2$ from above or below.

A word of caution is now in order.  Finite-$\varepsilon$ corrections
to the result in Eq.~(\ref{varvar}) are of order of
$\mathcal{O}(\varepsilon^{2 - \mu})$ for $1 < \mu < 2$. Therefore the
asymptotic behavior above can be only attained when $\varepsilon \ll
\exp\left(-1/(2 - \mu)\right)$.  In other words, the variance can be
made arbitrarily small by choosing $\mu$ closer to $2$, but only at
the expense of increasing the experimental resolution ($t_0 \to 0$ or
$T \to \infty$).

To confirm our analytical results we simulated random walks on a $3d$
lattice and computed $P(u_\mu)$ using Eq.~(\ref{u}) from a large
ensemble of trajectories, for different values of $\mu$ and different
resolution $\varepsilon$.  For $\mu<1.5$ or $\mu>2.5$, the variance
computed numerically is well described by Eq.~(\ref{varvar}) and is
independent of $\varepsilon$ (Fig.~\ref{fig1}).  Near $\mu=2$,
corrections due to the finite resolution are noticeable, but the
numerics clearly show that the variance of the distribution $P(u_\mu)$
decreases as $\varepsilon \rightarrow 0$.

\begin{figure}[ht]
  \psfrag{PU}[b][b][1]{$P(u_\mu)$}
  \psfrag{U}[b][b][1]{$u_\mu$}
  \psfrag{PUs}[b][b][0.9]{$P(u_\mu)$}
  \psfrag{Us}[b][b][0.9]{$u_\mu$}
  \centerline{\includegraphics*[width=0.5\textwidth]{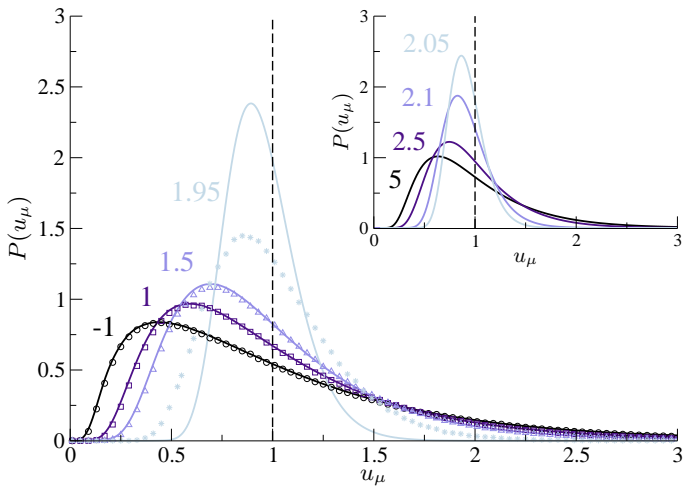}}
  \caption{(color online) The distribution $P(u_{\mu})$ for different
    $\mu \neq 2$ in 3D systems. The curves correspond to numerical
    inversion of Eq.~(\ref{alpha<2}) and the symbols to direct
    numerical simulations of random walks: from dark to light to $\mu
    = - 1$ (circles), $\mu = 1$ (squares), $\mu = 3/2$ (triangles),
    and $\mu = 1.95$ (stars). The numerical values were obtained for
    $\varepsilon=10^{-5}$, except for $\mu = 1.95$ for which we used
    $\varepsilon=10^{-7}$. Recall that $\mu = - 1$ corresponds to LSE
    and $\mu = 1$ to MLE \cite{berglund,michalet,mb,boyer,boyer1}.  In
    the inset we depict the curves corresponding to the inversion of
    Eq.~(\ref{alpha>2}): from dark to light $\mu = 5$, $2.5$, $2.1$
    and $2.05$.}
\label{fig2}
\end{figure}

The large- and small-$u$ asymptotics of $P(u_{\mu})$ can be deduced
directly from Eqs.~(\ref{alpha<2}) and (\ref{alpha>2}). For $\mu < 2$
and $u_{\mu} \ll 1$, $P(u_{\mu})$ shows a singular behavior:
\begin{equation}
\label{short}
P(u_{\mu}) \sim \exp\left(- \frac{d^2}{4 \chi_1 u_{\mu}}\right)
\frac{1}{u_{\mu}^{\zeta}}, \, \zeta = \frac{3}{2} + \frac{d}{4}
\frac{\mu}{|2 - \mu|}\,.
\end{equation}
The asymptotic behavior for $\mu > 2$ can be obtained from
Eq.~(\ref{short}) by simply replacing $\chi_1 \to \chi_2$.  Note that
Eq.~(\ref{short}) describes a bell-shaped function with a maximal
value $u^* \to 1$ when $\mu \to 2$ from above or below for arbitrary
$d$. Next, for $u_{\mu} \gg 1$ and $\mu < 2$, we find
\begin{equation}
\label{long}
P(u_{\mu}) \sim u_{\mu}^{d/2 - 1}
\exp\left(- \frac{\chi_1 \gamma_{1-\nu,1}^2}{4} u_{\mu}\right) \,,
\end{equation}
where $\gamma_{\nu,1}$ is the 1st zero of the Bessel function
$J_{\nu}(z)$ \cite{abramowitz}.  Results for $\mu > 2$ follow from
Eq.~(\ref{long}) via the replacements $\chi_1 \to \chi_2$ and
$\gamma_{1-\nu,1} \to \gamma_{-\nu,1}$.  As $\mu$ gradually approaches
$2$, the distribution becomes increasingly narrow: the left tails
vanish because of the divergence of the factor $1/|2-\mu|$ in the
exponential, while the right tails vanish because $|2 - \mu|
\gamma_{-\nu,1}^2$ and $|2 - \mu| \gamma_{1-\nu,1}^2$ diverge.

The distributions $P(u_{\mu})$, obtained by inverting
Eqs.~(\ref{alpha<2}) and (\ref{alpha>2}), are plotted in
Fig.~\ref{fig2}. Indeed, the maximal value $u^* \to 1$ when $\mu \to
2$ either from above or from below. Already for $\mu = 1.95$ (or $\mu
= 2.05$) we get the most probable value $u^* \approx 0.94$, which
outperforms the LSE ($u^* \approx 0.47$) and the MLE ($u^* \approx
0.6$).  For $\mu = 1.95$ the variance ${\rm Var}(u_{\mu }) \approx
0.032$, which is an order of magnitude less than the variances
observed for LSE ($= 0.5$) and the MLE ($\approx 0.33$).  Similarly to
Fig.~\ref{fig1}, finite-resolution corrections are negligible for
$\mu<3/2$, and $P(u_{\mu})$ is well described by Eq.~(\ref{alpha<2}).
For $\mu=1.95$ and finite resolution $\varepsilon=10^{-7}$, we obtain
a broader distribution and with a smaller $u^*$ than the corresponding
to Eq.~(\ref{alpha<2}) for infinite resolution.  However, note that
the most probable value of $P(u_{1.95})$ that we obtain at finite
resolution is $\approx 0.84$, which outperforms the LSE and MLE for
infinite resolution.

\begin{figure}[ht]
  \psfrag{PU}[b][b][1]{$P(u_2)$}
  \psfrag{U}[b][b][1]{$u_2$}
  \centerline{\includegraphics*[width=0.5\textwidth]{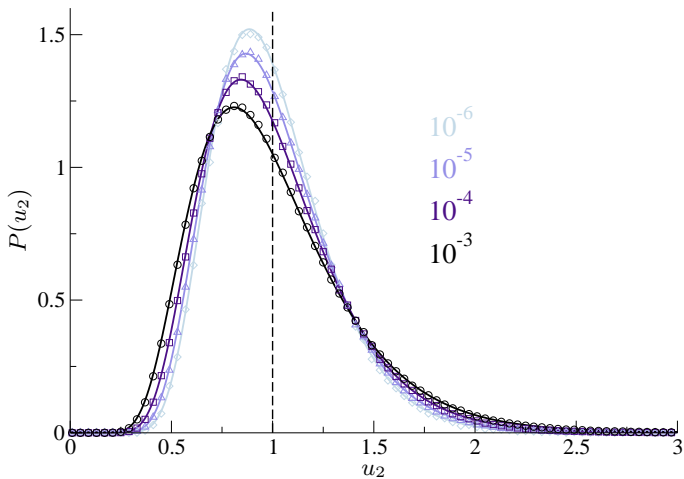}}
  \caption{(color online) The distribution $P(u_{2})$ obtained from a
    numerical inversion of Eq.~(\ref{12}) for 3D systems. The curves
    from the dark to light correspond to $\varepsilon =$ $10^{-3}$,
    $10^{-4}$, $10^{-5}$ and $10^{-6}$. The symbols are results of
    numerical simulations of random walks for $\varepsilon =$ $10^{-3}$
    (circles) and $\varepsilon =$ $10^{-4}$ (squares), $\varepsilon =$
    $10^{-5}$ (triangles) and $\varepsilon =$ $10^{-6}$ (diamonds).}
\label{fig3}
\end{figure}

We turn next to the case $\mu \equiv 2$ with $\varepsilon = t_0/T$
small but finite, seeking the variance and the distribution of $u_{\mu
  = 2}$. We consider a slightly more general form for $\omega(t)$:
\begin{equation}
\label{omlog}
\omega(t) =
\begin{cases} 2 \xi/t_0^2, & \text{for $t < t_0$,}
\\
1/t^2, &\text{for $t_0 \leq t \leq T$,}
\end{cases}
\end{equation}
where $\xi$ is a tunable amplitude. For such a choice, the moment
generating function is given explicitly by
\begin{equation}
\label{12}
\Phi(\sigma) = \left(\frac{2 \, \delta \,
\varepsilon^{(\delta-1)/2}}{\phi_{+}}\right)^{d/2} \left[1 + \frac{\phi_-}{\phi_+} \,
\varepsilon^{\delta}\right]^{-d/2} \,,
\end{equation}
\begin{equation}
\phi_{\pm} = \left(\delta \pm 1\right) \left({\rm ch}
\left(\sqrt{2 \gamma \xi \sigma}\right) \pm \frac{\delta \mp 1}
{2 \sqrt{2 \gamma \xi \sigma}} {\rm sh}\left(\sqrt{2 \gamma \xi \sigma}\right)
\right) \,, \nonumber\\
\end{equation}
where $\delta = \sqrt{1 + 4 \gamma \sigma}$ and $\gamma = 2/d (\xi +
\ln(1/\varepsilon))$. Differentiating Eq.~(\ref{12}), we find
\begin{equation}
\label{7}
{\rm Var}(u_2) = \frac{4}{3 d} \, \frac{3 \ln(1/\varepsilon) - 3 (1-\varepsilon) +
2 (1 - \varepsilon) \xi + \xi^2}{\left(\xi + \ln(1/\varepsilon)\right)^2} \,.
\end{equation}
It is a non-monotonic function of $\xi$ with a minimum at
\begin{equation}
\xi = \xi_{{\rm opt}} = \frac{(2 + \varepsilon) \ln(1/\varepsilon) - 3 (1 - \varepsilon)}
{\ln(1/\varepsilon) + \varepsilon - 1} \,.
\end{equation}
The corresponding  optimized variance is given by:
\begin{equation}
\label{optimal}
{\rm Var}_{\rm opt}(u_2) = \frac{4 }{3 d} \,
\frac{3 \ln(1/\varepsilon) - 4 + 5 \varepsilon - \varepsilon^2}{\ln(1/\varepsilon)
\left(\ln(1/\varepsilon) + 1 + 2 \varepsilon\right) - 3 (1 - \varepsilon)} \,.
\end{equation}
From Eq.~(\ref{optimal}) we find that in $3d$ ${\rm Var}_{\rm opt}(u_2)
\approx 0.144, 0.096, 0.082$ for $\varepsilon = 10^{-3}, 10^{-5},
10^{-6}$, respectively. When $\varepsilon \to 0$, ${\rm Var}_{\rm
  opt}(u_2)$ vanishes as
\begin{equation}
\label{asymptotic1}
{\rm Var}_{\rm opt}(u_2) \sim \frac{4}{d} \frac{1}{\ln(1/\varepsilon)}\,.
\end{equation}
Therefore, ${\rm Var}_{\rm opt}(u_2)$ can be made arbitrarily small
but at expense of a progressively higher resolution.  In the limit
$\varepsilon \to 0$ the distribution converges to a delta-function.
The estimators with $\mu = 2$ are the only, in the family defined by
Eq.~(\ref{u}), that possess an ergodic property.  This is shown in
Fig.~\ref{fig3} where we plot $P(u_{2})$ obtained by numerically
inverting Eq.~(\ref{12}) for different resolutions. The symbols in
this figure correspond to numerical simulations using the weight
function of Eq.~(\ref{omlog}).

Finally let us consider the case of BM recorded at discrete time steps
$\Delta \cdot j$, $j=1, \ldots ,N$, ($T = \Delta \cdot N$), which is
an important in its own right problem but also will allow us to
justify the choice of the weight function in Eq.~(\ref{weight1}).  We
focus on the estimator of a general form
\begin{equation}
\label{tilde}
\tilde{u} = \frac{1}{2 d \Delta} \sum_{j = 1}^N \omega_j {\bf B}^2_{\Delta \cdot j}/
\sum_{j = 1}^N j \, \omega_j \,,
\end{equation}
where $\omega_j$ now is an arbitrary weight function. The culminating
point of our analysis is to determine, via a variational approach, the
function $\omega_j$ which yields the lowest possible variance ${\rm
  Var}(\tilde{u})$, given from Eq.~(\ref{tilde}) by
\begin{equation}
\label{g}
{\rm Var}(\tilde{u}) = \frac{2}{d} \sum_{j = 1}^N \omega_j \, \sum_{k = 1}^N
\omega_k \min(k,j)^2\, /\left(\sum_{j = 1}^N j \, \omega_j\right)^2
\end{equation}
where $\min(k,j)$ is the minimum of  $k$ and $j$. Minimizing
\begin{equation}\label{func2}
\tilde{F} = \frac{1}{2} \sum_{j = 1}^N \omega_j \, \sum_{k = 1}^N \omega_k
\min(k,j)^2 - \lambda \left(\sum_{j = 1}^N j \, \omega_j - 1\right) \,,
\end{equation}
with respect to each $\omega_j$ ($\lambda$ is a Lagrange multiplier
enforcing the constraint $\mathbb{E}\left\{\tilde{u}\right\} = 1$), we
find that the {\em optimal} weight obeys
\begin{equation}
\sum_{j = 1}^N \omega_j \min(k,j)^2 = \lambda \cdot k \,, k=1, \ldots, N\,
\end{equation}
which can be solved exactly
to give
\begin{equation}
\lambda = N \left(\sum_{k = 1}^N \frac{k}{4 k^2 - 1}\right)^{-1}
\end{equation}
and
\begin{equation}
\label{weight}
\omega_j =
\frac{2 \lambda}{4 j^2 - 1} = \left(
\frac{N}{\sum_{k = 1}^N \frac{k}{4 k^2 - 1}}\right) \frac{1}{4 j^2 - 1} \,.
\end{equation}
The optimal variance in this case reads
\begin{equation}
\label{optimal2}
{\rm Var}(\tilde{u}) = \frac{1}{d}
\left(\sum_{j=1}^N \frac{j}{(4 j^2 - 1)}\right)^{-1} \, .
\end{equation}
Therefore, the weight function in Eq.~(\ref{weight}) minimizes the
discretized least squares functional in Eq.~(\ref{func2}) and produces
an ergodic estimator: the smallest possible variance (for the class of
estimators defined by Eq.~(\ref{tilde})) vanishes as $N\to \infty$.
Choosing some initial time lag and turning to the limit $\Delta \to 0$
and $N \to \infty$, while keeping $\Delta N = T$ fixed, the weight
function in Eq.~(\ref{weight}) converges to the form in
Eq.~(\ref{weight1}) with $\mu = 2$, which thus justifies our choice of
the power-law trial weight function for continuous-time Brownian
motion. Note that for $N \gg 1$, the leading asymptotic behavior of
the variance in Eq.~(\ref{optimal2}) coincides with
Eq.~(\ref{asymptotic1}), but produces slightly higher values of the
variance (as the former estimator is based on an everywhere discrete
process).

To conclude, we have analyzed the ergodic properties and the
asymptotic behavior of a family of least-squares estimators in
Eq.~(\ref{u}). We have demonstrated that the estimators with $\mu = 2$
are the only that possess an ergodic property, {\it i.e.}, they can
provide the true ensemble averaged diffusion coefficient from a single
trajectory data with any necessary precision, but at expense of a
progressively higher experimental resolution. Convergence to the true ensemble 
average value appears, however, to be rather slow: the variance of such an optimal estimator
vanishes only in proportion to $1/\ln(T)$. This means that for practical purposes 
the methods based  
on two-time correlation functions can provide better estimators, because the variance 
of the corresponding estimator decays faster, as $1/T$, even in the presence of localisation errors \cite{berglund,michalet,mb}.   


\begin{acknowledgments}
  We thank Gregory Putzel for valuable comments.  The authors
  acknowledge partial support from the European Science Foundation
  through the Research Network "Exploring the Physics of Small
  Devices". CMM is supported by the European Research Council and the
  Academy of Finland.
\end{acknowledgments}

\end{document}